\documentstyle[12pt,psfig]{article}
\begin{document}
\begin{center}

\baselineskip=24pt
{\Large \bf Characteristics of Alpha, Gamma and Nuclear Recoil Pulses from 
NaI(Tl) at 10-100 keV Relevant to Dark Matter Searches}

\vspace{1cm}

{\large V.A.Kudryavtsev$^a$, M.J.Lehner$^a$, C.D.Peak$^a$, 
T.B.Lawson$^a$, 
P.K.Lightfoot$^a$, J.E.McMillan$^a$, J.W.Roberts$^a$, 
N.J.C.Spooner$^a$, 
D.R.Tovey$^a$, C.K.Ward$^a$, P.F.Smith$^b$, N.J.T.Smith$^b$}
\baselineskip=18pt

\vspace{0.5cm}

\noindent $^a$ {\it Department of Physics and Astronomy, 
University of Sheffield, Sheffield S3 7RH, UK}

\noindent $^b$ {\it Rutherford Appleton Laboratory, Chilton, OX11 0QX, UK}

\vspace{1.0cm}

{\large \bf Abstract}

\end{center}

\indent Measurements of the shapes of scintillation pulses produced by 
nuclear recoils, alpha 
particles and photons in NaI(Tl) crystals at visible energies of 
10-100 keV have been performed 
in order to investigate possible sources of background in NaI(Tl) 
dark matter experiments and, 
in particular, the possible origin of the anomalous fast time 
constant events observed in the UK 
Dark Matter Collaboration experiments at Boulby mine \cite{Smith}. Pulses 
initiated by X-rays (via 
photoelectric effect close to the surface of the crystal) were found 
not to differ from those 
produced by high-energy photons (via Compton electrons inside the 
crystal) within 
experimental errors. However, pulses induced by alpha particles 
(degraded from an external 
MeV source) were found to be $\sim10\%$ faster than those of nuclear 
recoils, but insufficiently fast to account for the anomalous events.

\vspace{1.0cm}

\noindent PACS: 29.40Mc, 14.80 Ly

\noindent Keywords: Scintillation detectors; Dark matter; WIMP; 
Radioactive sources; Pulse-shape discrimination

\vspace{1cm}

\noindent Corresponding author: V. A. Kudryavtsev, Department of Physics and 
Astronomy, University 
of Sheffield, Hicks Building, Hounsfield Rd., Sheffield S3 7RH, UK

\noindent Tel: +44 (0)114 2224531; Fax: +44 (0)114 2728079

\noindent E-mail: v.kudryavtsev@sheffield.ac.uk

\pagebreak

\noindent {\large \bf 1. Introduction}
\vspace{0.5cm}

\indent Several NaI-based experiments searching for Weakly Interacting 
Massive Particles 
(WIMPs), as possible constituents of galactic dark matter, use 
pulse-shape discrimination to 
separate the nuclear recoil signals expected from WIMP elastic 
scattering from electron recoils 
due to gamma background (see for example \cite{Gerbier,Smithold,Bernabei}). 
The feasibility of this technique arises 
because pulses initiated by nuclear recoils in NaI(Tl) are known 
to be typically $30\%$ faster than 
those due to electron recoils \cite{Smithold,Tovey,Quenby}. 
In each case the integrated 
scintillation pulses can be 
adequately fitted by assuming an exponential decay of time 
constant $\tau$. The resulting distribution 
of $\tau$, can be approximated by a gaussian in $\ln(\tau)$. The shape of 
the pulses can then be 
characterised by the mean value of $\tau$, $\tau_o$, in the gaussian 
\cite{Smithold}. 
The difference between $\tau_o$ values 
for nuclear and electron recoils is of the order of $30\%$ for 
measured energies above 20 keV, 
decreasing almost to zero at 2 keV. This difference, divided by 
the distribution width, is a 
measure of the discrimination which can be improved by 
optimising the crystal operating temperature \cite{Smith}.

Assuming operation of the technique at energies higher 
than $\sim$4 keV \cite{Smith,Smithold} then, apart from 
gammas, other potential sources of background are neutrons 
and alpha particles. Neutron 
interactions are expected to yield events indistinguishable 
from WIMP interactions (hence 
neutrons are usually used to determine the response of NaI to 
nuclear recoils since neutron 
scattering on nuclei is similar to WIMP scattering in terms of 
nuclear recoil generation). 
However, neutron background can be sufficiently suppressed by 
shielding the detectors from 
neighbouring radioactivity and using deep underground sites 
to avoid neutrons from cosmic-ray 
muons. In the case of alpha particles any background at low 
energies cannot arise from the 
likely small contamination of uranium and thorium in NaI(Tl) 
because the alphas from uranium 
and thorium decay chains have energies exceeding 1 MeV. However, 
high energy alphas from 
activity in surrounding materials potentially may penetrate into 
the crystal and deposit energy in 
the surface layers. Their path in the sensitive surface layer 
of a crystal, and hence their energy 
deposition, can then be small depending on the point of their 
production. Thus such alphas 
could be responsible for background events and may be responsible 
for the anomalous fast time 
constant events (pulses faster than recoil-like pulses) observed 
in UKDMC NaI(Tl) experiments undertaken at Boulby mine 
\cite{Smith,Spooner}.

In this paper we report characterisation of the shape of pulses 
induced by neutrons, alphas 
and electron recoils in NaI(Tl) in order to investigate possible 
sources of background in NaI(Tl) 
dark matter experiments. Similar measurements have been 
performed previously in NaI(Tl) 
(see, for example, \cite{Gerbier,Smithold,Tovey} and references therein). 
However, presented here are results obtained 
with a single NaI(Tl) crystal at low energies (10-100 keV), 
allowing direct comparison of the 
pulses induced by the different particles in the energy 
range of prime interest for dark matter experiments.

\vspace{0.8cm}
\noindent {\large \bf 2. Experimental set-up and analysis technique}
\vspace{0.5cm}

\indent Measurements were performed using a Hilger Analytical Ltd. 
unencapsulated NaI(Tl) 
crystal of dimensions 2 inch x 2 inch diameter cooled, using 
a purpose-built copper cryostat, to 
11$^o$C with control so that variation did not exceed ±0.5$^o$C.  
Further details of the apparatus can 
be found in \cite{Toveythesis,Toveynew}. The temperature chosen was close 
to that of the crystals used in the UK Dark 
Matter Collaboration (UKDMC) experiments at Boulby mine \cite{Smith,Spooner}. 
The crystal was viewed 
through a pair of silica light guides by two 3-inch ETL type 
9265 photomultiplier tubes (PMTs). 
Integrated pulses from the PMTs were digitised using a LeCroy 
9430 oscilloscope driven by a 
Macintosh computer running Labview-based data acquisition software 
identical to that used with 
the UKDMC dark matter detectors.  The digitised pulse shapes 
were passed to the computer and 
stored on disk. For the final analysis the sum of the pulses 
from two PMTs was used. The 
apparatus was found to yield a total light output corresponded 
to $\sim$5.5 photoelectrons/keV.

Energy calibrations for the tests were performed with a 
$^{57}$Co gamma source (122 keV).  A 
$^{60}$Co gamma source was used to obtain pulse shapes from 
high-energy gammas. These gammas 
undergo Compton scattering inside the crystal producing electrons 
which deposit energy 
measured by the detector.  To measure the shapes of the 
integrated pulses induced by nuclear 
recoils (sodium and iodine), the crystal was irradiated by 
neutrons from a $^{252}$Cf source. Neutron 
energies were decreased by shielding the source with a 
4 cm thick lead block. All 
aforementioned sources were placed outside the copper vessel 
containing the crystal, light 
guides and PMTs. An $^{241}$Am source was used to irradiate the 
crystal with alpha particles. 
However, since the path length of alphas does not exceed tens 
of microns the source was 
attached to the bare cylindrical face of the crystal inside 
the cryostat. To decrease the energy of 
the alphas down to the keV range of interest a few, thin ($\sim$10 $\mu$m), 
layers of plastic were placed 
between the source and the crystal. The 60 keV gamma-line 
from the $^{241}$Am source also 
allowed independent energy calibration of the detector and 
measurement of the shape of the 
pulses initiated by electrons near the crystal surface. 
Such electrons are produced via the photoelectric effect.
	
Analysis was performed, using the procedure described 
in \cite{Smithold}, by fitting an exponential to 
each integrated pulse to obtain the index of the exponent, 
$\tau$ - faster pulses (due to nuclear 
recoils, for example) having smaller values of $\tau$, than slower 
pulses such as from gammas. For 
each experiment histograms were generated of the number of 
detected events versus the value of 
$\tau$ - referred to here as $\tau$-distributions. The $\tau$-distributions 
are known to be dependent on the 
measured energy. To reduce such dependence the energy range 
(0-100 keV) was subdivided 
into energy bins of 10 keV width. We note that the energy 
threshold was about 10 keV. As 
shown in \cite{Smithold} (see also references therein) the $\tau$-distribution 
for each population of pulses can be 
approximated by a gaussian in $\ln(\tau)$ (for a more detailed 
discussion of the distributions see \cite{Toveythesis} and references therein):

\begin{equation}
{{dN}\over{d\tau}} = {{N_o}\over{\tau \sqrt{2\pi} \ln w}} \cdot
\exp \Big[ {{-(\ln \tau-\ln \tau_o)^2}
\over{2(\ln w)^2}} \Big]
\end{equation}
	 					
The $\tau$-distributions were fitted with a gaussian in $\ln(\tau)$
with several free parameters as 
follows. In the case of events from the $^{60}$Co gamma source a 
3-parameter fit was used with free 
parameters $\tau_o$, $w$  and $N_o$. In the experiments with the $^{252}$Cf 
neutron source both neutrons and 
gammas (from the source as well as from local radioactivity) 
were detected. The resulting $\tau$-
distribution can thus be fitted with two gaussians. However, 
the parameters $\tau_{o\gamma}$ and $w$ for 
events initiated by gammas (effectively by Compton electrons) 
are known from experiments 
with the $^{60}$Co gamma source. Assuming the value of $w$ (called 
the width parameter)  for the 
neutron distribution (where the pulses are due to nuclear 
recoils) is the same as that of the 
gamma distribution, since the width is determined mainly by 
the number of photoelectrons, 
again a 3-parameter fit can be applied.  In this case the free 
parameters are: number of neutrons, 
$N_{on}$, number of gammas, $N_{o\gamma}$, and the mean value of the exponent 
for the neutron distribution, 
$\tau_{on}$. In practice, for direct comparison of the gamma distributions 
obtained with different 
sources, the $\tau$-distribution of gamma events measured with the 
gamma source was used, instead 
of the gaussian fit, to approximate the distribution of gamma 
events measured with the neutron source.

In experiments with the alpha source both alphas and gammas 
were detected. To 
approximate the resulting distribution, we again used the 
$\tau$-distribution of the gamma events 
measured with $^{60}$Co gamma source and a gaussian fit to the 
alpha distribution with 3 free 
parameters: number of alphas, $N_{o\alpha}$, number of gammas, $N_{o\gamma}$, 
and the mean value of the 
exponent of the alpha distribution, $\tau_{o\alpha}$. Furthermore, by 
making use of the 60 keV X-rays from 
the $^{241}$Am alpha source it was also possible to evaluate $\tau_{oX}$, 
attributed to pulses due to X-ray 
events initiated via the photoelectric effect near the surface 
of the crystal.  Finally, to compare 
the 4 populations of events (initiated by gammas, neutrons, 
alphas and X-rays), we compared 
the values of $\tau_{o\gamma}$, $\tau_{on}$, $\tau_{o\alpha}$ and $\tau_{oX}$.

\vspace{0.8cm}
\noindent {\large \bf 3. Results and discussion}
\vspace{0.5cm}

\indent Measured $\tau$-distributions for events in two example energy 
bins are plotted in Figure 1a 
(30-40 keV, alpha source), 1b (55-65 keV, alpha source), 
1c (30-40 keV, neutron source) and 
1d (55-65 keV, neutron source). Plus signs show the data 
collected with the aforementioned 
sources. Open squares correspond to the data collected 
with the $^{60}$Co gamma source normalised 
using the best fit procedure. Dotted curves show the fits 
to the neutron (alpha) distributions. 

Data collected with the gamma source (squares in Figure 1) 
match well the right-hand 
parts of distributions obtained with the neutron or alpha 
sources (these parts correspond to 
gamma events detected in the experiments with the neutron 
or alpha sources). This is true also 
for the 55-65 keV range with the alpha source where the gamma 
(right-hand) part of the $\tau$-distribution 
is dominated by X-rays from the $^{241}$Am source. 
The amplitude of the right-hand 
peak at 55-65 keV (Figure 1b) is several times more than 
that at 30-40 keV (Figure 1a) (note the 
logarithmic scale of the $y$-axes), showing the presence of 
the strong 60 keV line superimposed 
on the background due to Compton electrons. This means that 
the $\tau$-distribution, and hence the 
basic shape of the pulses due to the X-rays, does not differ 
from that of the Compton electrons 
initiated by high-energy gammas. It is clear also that the 
positions of the left-hand peaks in the 
experiment with the alpha source (Figures 1a and 1b) are 
shifted to the left with respect to the 
positions of the left-hand peaks in the experiment with 
the neutron source (Figures 1c and 1d, 
respectively). This is an indication that $\tau_{o\alpha}$ is less 
than $\tau_{on}$ and, hence, the pulses due to alphas 
are faster than the pulses due to nuclear recoils (note 
that it was shown in \cite{Tovey,Toveythesis} that pulses due 
to sodium recoils are indistinguishable from those initiated 
by iodine recoils at all energies of interest).
 
Figure 2 shows the fits to the $\tau$-distributions from the 
gamma- (solid curve), neutron- 
(dashed curve) and alpha- (dotted curve) induced events 
for the energy bin 30-40 keV. The total 
number of events in each case is normalised to unity.

The results are summarised quantitatively in terms of $\tau_o$  
in Table 1. The typical error in $\tau_o$ 
is of the order of 2-5 ns (arising from the statistics of 
the fit), except for the first energy bin in 
the experiment with the alpha source where the error of $\tau_{o\alpha}$ 
is 15 ns - being higher due to the 
smaller number of detected alphas. The value of $\tau_{oX}$, obtained 
from the fit to the right-hand peak 
of the $\tau$-distribution at 55-65 keV with the alpha source (see 
Figure 1), is 322$\pm$2 ns, in good 
agreement with the value of $\tau_{o\gamma}$ (2nd column of Table 1). 
However, this does not agree with the 
conclusion of \cite{Gerbiernew}, where the shapes of the pulses due to 
X-rays from an $^{241}$Am source and 
Compton electrons from high-energy gammas were found to be 
different (the shape of the 
pulses due to X-rays was found to be similar to that of 
nuclear recoils).  The values of $\tau_{o\alpha}$ (4th 
column of Table 1) are on average $10\%$ smaller than those 
of $\tau_{on}$ (3rd column of Table 1).

The values of $\tau_o$ are known to vary from one crystal 
to another depending on the growth 
technology, Tl doping, temperature and other factors 
\cite{Smith,Gerbier,Smithold,Bernabei,Tovey}. 
However, the ratios, for instance of 
$\tau_{on}$ to $\tau_{o\gamma}$ are known to be quasi-independent of the 
crystal for fixed energy and temperature (we 
found it to decrease from 0.80 down to 0.76 with increasing 
energy from 10 to 80 keV in the 
crystal with anomalous events, currently under operation in 
the Boulby mine). The ratios $\tau_{on}/\tau_{o\gamma}$, 
$\tau_{o\alpha}/\tau_{o\gamma}$ and $\tau_{o\alpha}/\tau_{on}$ are shown 
in the 5th, 6th and 7th columns of Table 1, respectively. The first 
two slightly decrease with increasing energy, while $\tau_{o\alpha}/\tau_{on}$ 
remains almost constant. The average 
ratio is $<\tau_{o\alpha}/\tau_{on}>$ = $0.90\pm0.01$. The ratio 
$<\tau_{o\alpha}/\tau_{on}>$ is higher 
than the ratio $<\tau_{oa}/\tau_{on}>$ = $0.79\pm0.04$ 
found for the anomalous fast events (pulses faster than 
recoil-like pulses) observed in the 
UKDMC experiment at Boulby mine \cite{Smith,Spooner}. This suggest that 
the anomalous events are not 
produced by external high energy alphas degraded in energy 
by a non-scintillating layer of 
material, assuming the ratio $<\tau_{o\alpha}/\tau_{on}>$ does not depend 
on the crystal.

\vspace{0.8cm}
\noindent {\large \bf 4. Conclusions}
\vspace{0.5cm}

\indent The form of the pulses initiated by gammas, alphas, nuclear 
recoils and X-rays have been 
analysed in terms of the mean value, $\tau_o$, of the gaussian 
distribution of exponent indices (see eq. 
(1)). The value of $\tau_{oX}$ of events initiated by X-rays (using 
the 60 keV line from an $^{241}$Am source) 
was found to be the same as that of events induced by 
Compton electrons from high-energy 
gammas. The values of $\tau_{o\alpha}$ for alpha events are smaller 
(by $\sim10\%$ on average) than those of $\tau_{on}$ 
for nuclear recoils induced by neutrons. However, the 
ratio of $\tau_{o\alpha}/\tau_{on}$ is higher than the 
corresponding ratio for the anomalous events to nuclear recoil 
events observed in the UKDMC 
experiment.  This suggests that the anomalous events are not 
produced by external high energy 
alphas degraded in energy by a non-scintillating layer of 
material, assuming the ratio $\tau_{o\alpha}/\tau_{on}$ does 
not depend on the crystal. 

\vspace{0.8cm}
\noindent {\large \bf 5. Acknowledgements}
\vspace{0.5cm}

The authors wish to thank for the support PPARC, Zinsser Analytic 
(J.E.M.), Hilger 
Analytical Ltd. (J.W.R.), Electron Tubes Ltd. (J.W.R.).

\pagebreak

\begin{table}[htb]
\caption{ Mean values, $\tau_o$, of gaussian distributions (see eq. (1)) 
for gamma- (column 2), 
neutron- (column 3) and alpha- (column 4) induced events, and 
ratios of the mean values 
(columns 5, 6 and 7). The typical errors are 2-5 ns (in $\tau_o$, 
see text for details) and 0.02 (in the ratios).}
\vspace{1cm}
\begin{center}
\begin{tabular}{|c|c|c|c|c|c|c|}\hline
Energy, keV & $\tau_{o\gamma}$, ns & $\tau_{on}$, ns & $\tau_{o\alpha}$, ns & 
$\tau_{on}/\tau_{o\gamma}$ & $\tau_{o\alpha}/\tau_{o\gamma}$ &
$\tau_{o\alpha}/\tau_{on}$ \\ \hline
 10-20 & 292 & 242 & 220 & 0.83 & 0.75 & 0.91 \\ \hline
 20-30 & 307 & 242 & 220 & 0.79 & 0.72 & 0.91 \\ \hline
 30-40 & 314 & 241 & 219 & 0.77 & 0.70 & 0.91 \\ \hline
 40-50 & 318 & 240 & 218 & 0.75 & 0.69 & 0.91 \\ \hline
 50-60 & 320 & 240 & 214 & 0.75 & 0.67 & 0.89 \\ \hline
 60-70 & 320 & 240 & 214 & 0.75 & 0.67 & 0.89 \\ \hline
 70-80 & 321 & 240 & 215 & 0.75 & 0.67 & 0.90 \\ \hline
 80-90 & 322 & 239 & 216 & 0.74 & 0.67 & 0.90 \\ \hline
90-100 & 322 & 238 & 216 & 0.74 & 0.67 & 0.91 \\ \hline
\end{tabular}
\end{center}
\end{table}

\pagebreak
 
\begin{figure}[htb]
\begin{center}
\psfig{figure=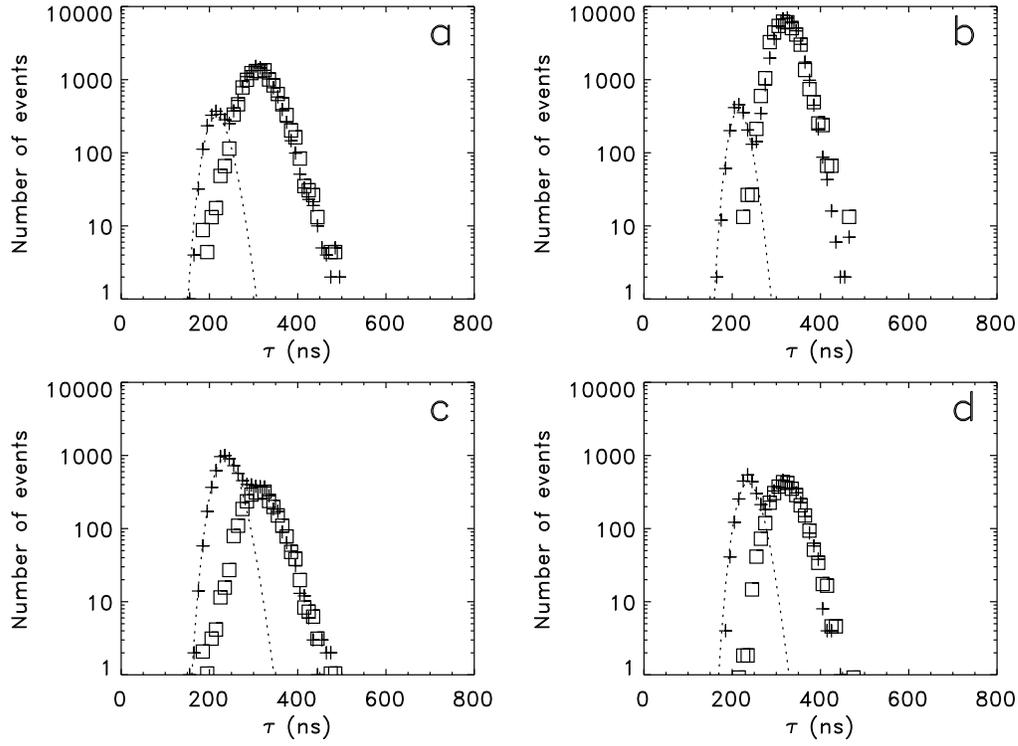,height=10cm}
\caption{ $\tau$-distributions for events collected with different 
sources: a) alpha source, measured 
energy 30-40 keV; b) alpha source, measured energy 55-65 keV; 
c) neutron source, measured 
energy 30-40 keV; d) neutron source, measured energy 55-65 keV. 
Plus signs show the data 
collected with the aforementioned sources. Open squares correspond to 
the measured 
distribution of Compton electrons (from $^{60}$Co events) 
normalised using the best fit procedure. 
Dotted curves are the best fits to the left-hand parts of 
the distributions (alpha or neutron induced events).}
\end{center}
\end{figure}

\begin{figure}[htb]
\begin{center}
\psfig{figure=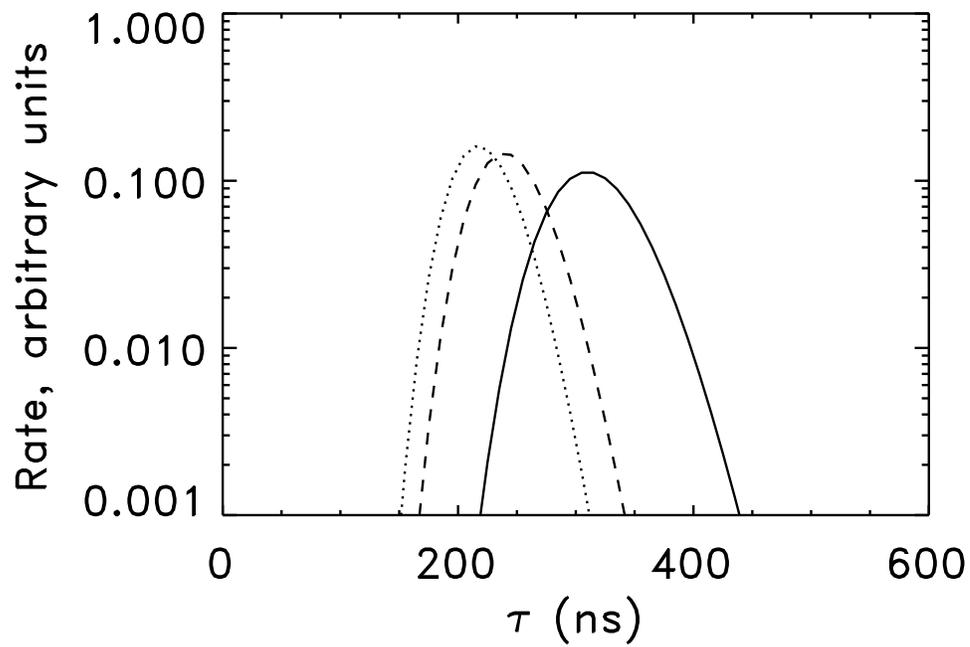,height=10cm}
\caption{ Fits to $\tau$-distributions for gamma- (solid curve), 
neutron- (dashed curve) and alpha- 
(dotted curve) induced events for the energy bin 30-40 keV. 
The total number of events for each fit is normalised to unity.}
\end{center}
\end{figure}

\end{document}